\documentclass [12pt,twoside]{article}           
\usepackage[left,modulo]{lineno}
\usepackage{setspace}

\countdef\arxiv=201

\ifnum\arxiv=0 \input cpreb.tex \fi

\ifnum\arxiv=1

\usepackage{epsfig,times,lscape}
\usepackage[usenames]{color}

    \ifnum\count102=1

    \fi

    \ifnum\count102=2
\fi

\newcommand{\nc}{\newcommand}

     \ifnum\count101=1
\nc{\qI}[1]{\section{{#1}}}
\nc{\qA}[1]{\subsection{{#1}}}
\nc{\qun}[1]{\subsubsection{{#1}}}
\nc{\qa}[1]{\paragraph{{#1}}}

\def\qpar{\vskip 2mm plus 0.2mm minus 0.2mm}
\def\qL{\hfill \break}
     \fi 

      \ifnum\count101=2
 \nc{\qI}[1]{\parindent=0mm \vskip 8mm 
{\centerline{\LARGE \color{red}#1}}\vskip 3mm}
\nc{\qA}[1]{\vskip 2.5mm \noindent 
{{\bf\large\color{blue}  #1}} \vskip 1mm \parindent=0mm}
 \nc{\qun}[1]{\vskip 1mm \noindent {\sl #1 }\quad }

\def\qL{\hfill \break}
\def\qpar{\vskip 2mm plus 0.2mm minus 0.2mm}

      \fi

\def\qtb{\vrule height 0pt depth 5pt width 0pt}

\nc{\qfoot}[1]{\footnote{{#1}}}

      \ifnum\count101=1
\def\qbu{\hfill \par \hskip 6mm $ \bullet $ \hskip 2mm}
\def\qee#1{\hfill \par \hskip 6mm (#1) \hskip 2 mm}
      \fi
      \ifnum\count101=2
\def\qbu{\hfill \par \hskip 4mm $ \bullet $ \hskip 2mm}
\def\qee#1{\hfill \par \hskip 4mm (#1) \hskip 2 mm}
      \fi

\def\qparr{ \vskip 1.0mm plus 0.2mm minus 0.2mm \hangindent=10mm
\hangafter=1}

     \ifnum\count101=1 
 \def\qdec#1{\parindent=0mm\par {\leftskip=2cm {#1} \par}}
     \fi
     \ifnum\count101=2
  \def\qdec#1{\parindent=0mm \par {\leftskip=1cm {#1} \par}}
  
  \def\qcitb#1{\noindent \hbox to 102mm{\hfill \small #1} \vskip 1mm}
      \fi

 \def\qpages#1{\count102=0{\loop\advance\count102 by 1
 \null \vfill\eject \ifnum\count102<#1 \repeat}}

\def\qn#1{\eqno \hbox{(#1)}}

\def\qtb{\vrule height 0pt depth 5pt width 0pt}

\def\qv{\vskip 0.1mm plus 0.05mm minus 0.05mm}

\def\qhw{\hskip 1.5mm}
\def\qleg#1#2#3{\noindent {\bf \small #1\qhw}{\small #2\qhw}{\it \small #3}\qv }
\fi

\begin{document}
\thispagestyle{empty}



\markboth{{\sl \hfill  \hfill \protect\phantom{3}}}
        {{\protect\phantom{3}\sl \hfill  \hfill}}

\large

\color{yellow} 
\hrule height 20mm depth 10mm width 170mm 
\color{black}
\vskip -2.2cm 

 \centerline{\bf \Large The detection of cheating in }
\vskip 2mm
 \centerline{\bf \Large multiple choice examinations}
\vskip 8mm
\centerline{\large 
Peter Richmond$ ^1 $ and Bertrand M. Roehner$ ^2 $
}

\vskip 8mm

{\bf Abstract}\quad
Cheating in examinations is acknowledged by an increasing number of
organizations to be widespread. We examine two different approaches to
assess their effectiveness at detecting anomalous results, suggestive
of collusion, using data taken from a number of multiple-choice
examinations organized by the UK Radio Communication Foundation.
Analysis of student pair overlaps of correct answers is
shown to give results consistent with more orthodox statistical
correlations for which confidence limits as opposed to the less
familiar ``Bonferroni method'' can be used.
A simulation approach is also
developed which confirms the interpretation of the empirical approach.

\vskip 3mm
\centerline{\it Version of 30 March 2015}

\vskip 3mm
{\normalsize Key-words: Examination, cheating, detection, binomial
  trial, multinomial trial, dependent variables}
\vskip 2mm

\vskip 15mm

{\normalsize 
1: School of Physics, Trinity College Dublin, Ireland.
Email: peter\_richmond@ymail.com \qL
2: Institute for Theoretical and High Energy Physics (LPTHE),
University Pierre and Marie Curie, Paris, France. 
Email: roehner@lpthe.jussieu.fr
}

\vfill\eject

\qI{Background}

Cheating in examinations is now acknowledged by an increasing number
of academic institutions to be widespread. Wesolowsky (2000)
notes one 2005 US study (McCabe 2005) 
which reported that from 10,000 faculty, 44\% were aware
of student cheating in the previous 3 years yet never reported the
fact. He also points to a second study by the Josephson Institute of
Ethics also in USA. This surveyed 43,000 high school students in
October 2010, which found that more than one half
admitted cheating on test
during past 12 months and one third admitted cheating more than
twice.
Cizek (1999)
as a result of a survey of studies prior to 1999 concluded that
``cheating is rampant''. Most research since suggests cheating is more
widespread than is usually believed.  In the UK cheating has been
reported to occur during medical examinations (McManus 2005). 
But other specialisms
are not exempt. Nor is it confined to schools and universities. Many
professionals, aspiring car drivers, nuclear missile control
supervisors and amateur radio operators are all required to take
examinations in order to gain an appropriate license. It is also not
unknown for the teacher to collude with examinees. This was
illustrated by a recent article in Time Magazine (Docktorman 2014)
which reported the
indictment of a college Principal and four teachers for conspiracy and
provision of examination answers to students before the examination.
Multiple-choice examinations form a significant element of the testing
process in all these cases. Here we report the development and
implementation of a statistical method aimed at detecting cheating in
multiple choice amateur radio examinations.  Our empirical study is
complemented by a set of simulations, which offer additional insight
into the method and outcomes. 
\qpar
The next section outlines the way amateur radio examinations in the UK
are administered and managed together with initial approaches to
cheating detection.  A statistical approach to detection is then
described together with some output for three different examinations
that illustrate the potential of the method.  Finally we end with
thoughts and conclusions. 

Amateur radio has its origin in the early studies by Marconi and other
physicists of electromagnetic wave propagation at the end of the 19th
and beginning of the 20th centuries.  As a hobby it began to grow
substantially after world war one. It has been estimated that around
two million people world wide are regularly involved with amateur
radio.  In the UK for the past few years new entrants number about
1,000 annually. To participate, entrants across the world must obtain a
license. Most advanced countries issue this following success in a
national examination. In the UK, the government office of
communications regulation (OFCOM) confers responsibility for the
examination to the Radio Communications Foundation (RCF) an
independent charity established to support people and projects where
radio communication, and amateur radio in particular, is the
theme. The examination itself provides for three levels of license:
Foundation, Intermediate and Advanced. Here we are concerned with the
Advanced examination which at the moment consists of 62 multiple
choice questions that test knowledge of various aspects of radio
including the license conditions, basic electronics, radio and
transmitter architecture, antennae and electromagnetic wave
propagation, testing methods and safety. Each question is constructed
with the correct answer and three distractors. The examination is held
simultaneously at a number of approved centres located across the
UK. The number of candidates at any centre varies depending on demand
(see schematic shown in figure 1). As an example, in 2008 63
candidates sat for an advanced examination across 25 centres. The
average number of candidates per centre varies from 1 to 6.

\begin{figure}[htb]
\centerline{\psfig{width=12cm,figure=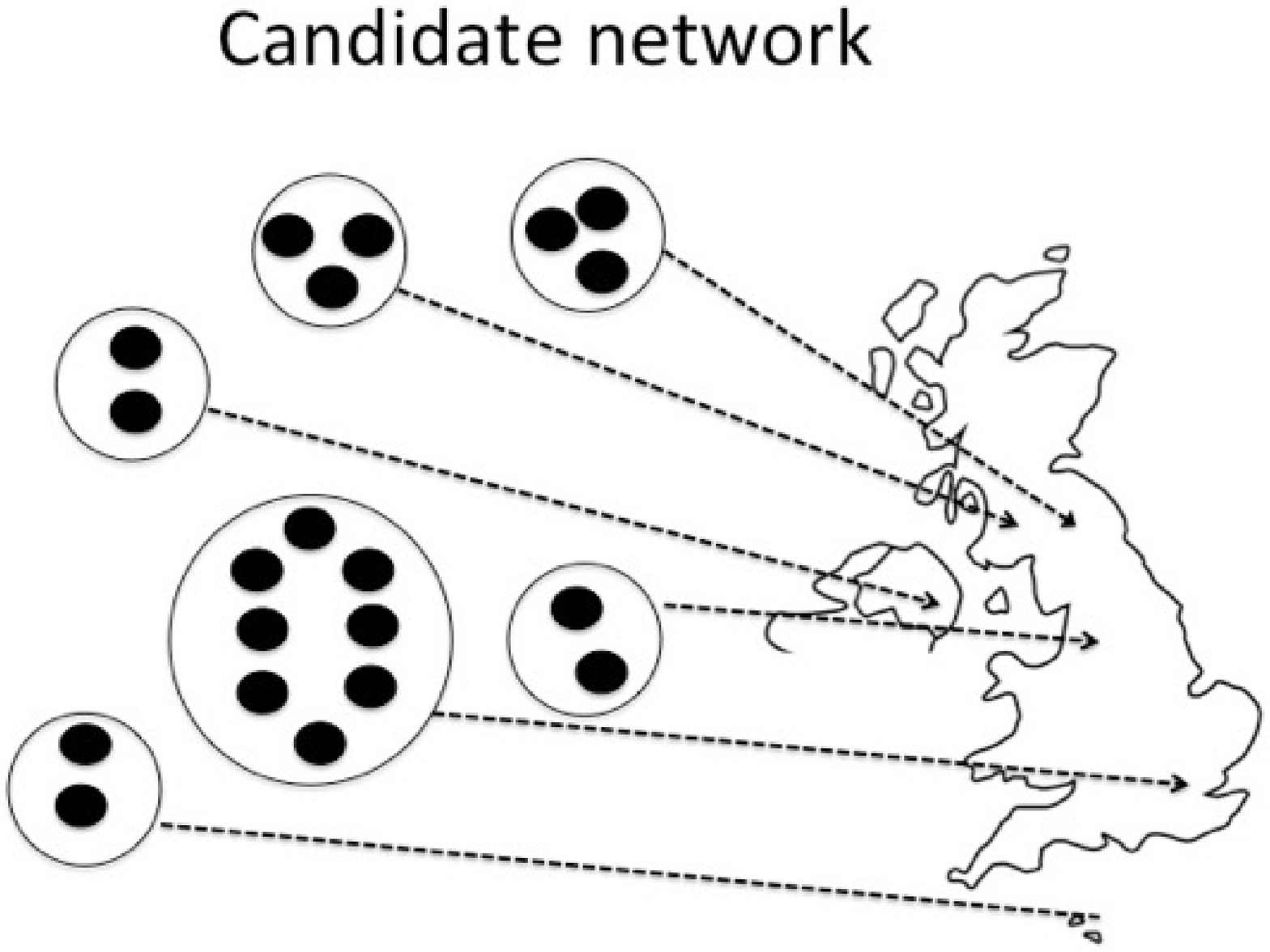}}
\qleg{Fig. 1: }
{Schematic illustration of the organization of examinations across the
  UK with relatively small numbers of students taking the examination
  at different centres simultaneously.}
{Source: }
\end{figure}

Locations are vetted and the conduct of the examination follows the
traditional pattern found in most schools and academic
establishments. Invigilators are required to be named and deemed
suitable by the RFC prior to the examination. However unlike schools,
the invigilators are volunteers and unpaid. Optical mark sheets record
the answers and for each candidate one has therefore a 62 character
string, viz: $ ABDC\ldots BCADC $ plus two additional identifiers for the
candidate and the particular centre in which the candidate sat the
examination.
\qpar
It can be noted that
for confidentiality reasons all identification code numbers of
candidates and examination centres have been removed from the
results given in the paper.

\qI{Analysis of empirical data}

\begin{table}[htb]

\centerline{\bf \small Table 1\quad Useful definitions for the
analysis of multiple choice examinations}

\vskip 5mm
\hrule
\vskip 0.5mm
\hrule

\color{black} 
\small

$$ \matrix{
\hbox{Correct answers}\hfill&4&1&4&4&3&3&3&4&1&4& & \cr
\hbox{} \hfill& & & & & & & & & & & \quad & \cr
\hbox{Answers of candidate 1 }(R_1)\hfill&4&3&4&4&4&1&1&4&1&2& & \cr
\hbox{Good (1)/wrong (0) answers of candidate 1
}(S_1)\hfill&1&0&1&1&0&0&0&1&1&0& & \cr
\hbox{Score of candidate 1 }(A_1) \hfill& & & & & & & & & & & \quad & 
\hbox{\bf 5}\cr
\hbox{} \hfill& & & & & & & & & & & \quad & \cr
\hbox{Answers of candidate 2 }(R_2) \hfill&4&1&4&4&3&4&3&4&1&4& & \cr
\hbox{Good (1)/wrong (0) answers of candidate 2
}(S_2)\hfill&1&1&1&1&1&0&1&1&1&1& &\cr
\hbox{Score of candidate 2 }(A_2) \hfill& & & & & & & & & & & \quad & 
\hbox{\bf 9}\cr
\hbox{} \hfill& & & & & & & & & & & \quad & \cr
\hbox{Overlap of correct answers }(A_{12})\hfill&&&&&&&&&&& &\hbox{\bf
  5}\cr
\hbox{Overlap of all answers }(T_{12})\hfill&&&&&&&&&&& &\hbox{\bf
  6}\cr
\hbox{} \hfill& & & & & & & & & & & \quad & \cr
(A_1+A_2)/2 \quad \hbox{(arithmetic mean score of the pair
  1,2)}\hfill&&&&&&&&&&& &\hbox{\bf 5.50}\cr
\sqrt{A_1A_2} \quad \hbox{(geometric mean score of the pair
  1,2)}\hfill&&&&&&&&&&& &\hbox{\bf 5.48}\cr
\hbox{Correlation between } R_1,R_2\hfill&&&&&&&&&&& &\hbox{\bf 0.35}\cr
\qtb
\hbox{Correlation between } S_1,S_2\hfill&&&&&&&&&&& &\hbox{\bf 0.33}\cr
\noalign{\hrule}
} $$
\vskip 0.5mm
Notes: For each question, the candidates had to select one of the 4 
proposed answers. In the examination there were 62 questions
but this table is limited to the first 10 answers. 
Candidates 1 and 2 were the first two candidates in the 
data set of 2008. They took the examination
in the same centre.
Altogether for the 19 centres there were 110 candidates.\qL
Initially, in the file of the UK Radio Communication
Foundation, the answers were coded in the form $ A,B,C,D $.
Thus, the first step was to
replace $ A,B,C,D $  by a numerical coding such 
as $ 1,2,3,4 $, which can be done either with a text editor or
with the Unix transliteration ``tr'' command. 
It can be seen that the arithmetic mean and the geometric mean
(used by McManus 2005) are very much the same. In following 
sections their common value will be referred to as the average
score of a pair of candidates.
\qL
{\it Source: Radio Communications Foundation of the United
Kingdom.}
\vskip 2mm
\hrule
\vskip 0.7mm
\hrule
\end{table}

In studying the data we follow a method similar to that proposed by
McManus et al. (2005)  who examined 
the overlap, $ A_{ij} $ of correct answers for
each and every pair of candidates in their cohort as a function of the
geometric mean of the scores, $ A_i $ and $ A_j $ for the pair of
students. 
In
our case it is useful to not only consider the total network of all
student pairs but also the sub-network formed by pairs of students
which sat the examination in each separate centre.  A feature of the
scatter plot such as is shown in figure 2 is that each data point
falls entirely inside the limited area defined by the dotted
triangle. This may be understood when one realizes that the maximum
value of the overlap occurs when the scores for each student are
equal; the minimum value of the overlap depends on the values
scored. If the students score less than 50\%, the minimum possible
value of the overlap is zero. For
students scoring over 50\% the minimum
overlap increases linearly from zero
to the maximum value when each
student scores 100\%.
\qpar

Assuming no contact by phone, internet or other wireless means during
the examination, we may expect the total overlap scores for the total
student network to be completely uncorrelated since it is
dominated by pairs from different examination centres.  This is shown
clearly using a dataset from an examination taken in 2012 and sat by a
total of 62 students equating to 1953 student pairs. \qL
Figure 2 shows
the outcome for the overlap $ A_{ij} $ plotted as a function of the square
root of the product of individual scores $ A_i $ and $ A_j $. 
The data points
for the total network fall about a mean, 
$ m = m\left(\sqrt{A_iA_j}\right) $  
that increases
as the geometric mean itself increases. The solid line is a one
parameter power law fit of this ``mean'' function. 
We deduce 
$  m\left(\sqrt{A_iA_j}\right)=62\left(\sqrt{A_iA_j}\right)^{\alpha} $
where the
exponent $ \alpha $ is chosen 
such that the sum of the overlap fluctuations
about the mean line is zero. In this instance $ \alpha=1.75 $.  The
distribution of these overlap fluctuations about the mean line fits
well a Gaussian distribution. Figure 3 shows the cumulative
distribution highlighting the final few points in the positive
tail. No unusual data outside the Gaussian distribution is apparent.
\qpar
We may now go further and examine the smaller sub-network of student
overlaps for only the students within centres. The small circles in
Figure 2 display the results for this sub-network. We observe no
unusual clustering between these data. Each point appears to be
distributed throughout the entire dataset. This suggests that the
argument often put forward that students who work together will score
similarly is not well founded.

\begin{figure}[htb]
\centerline{\psfig{width=12cm,figure=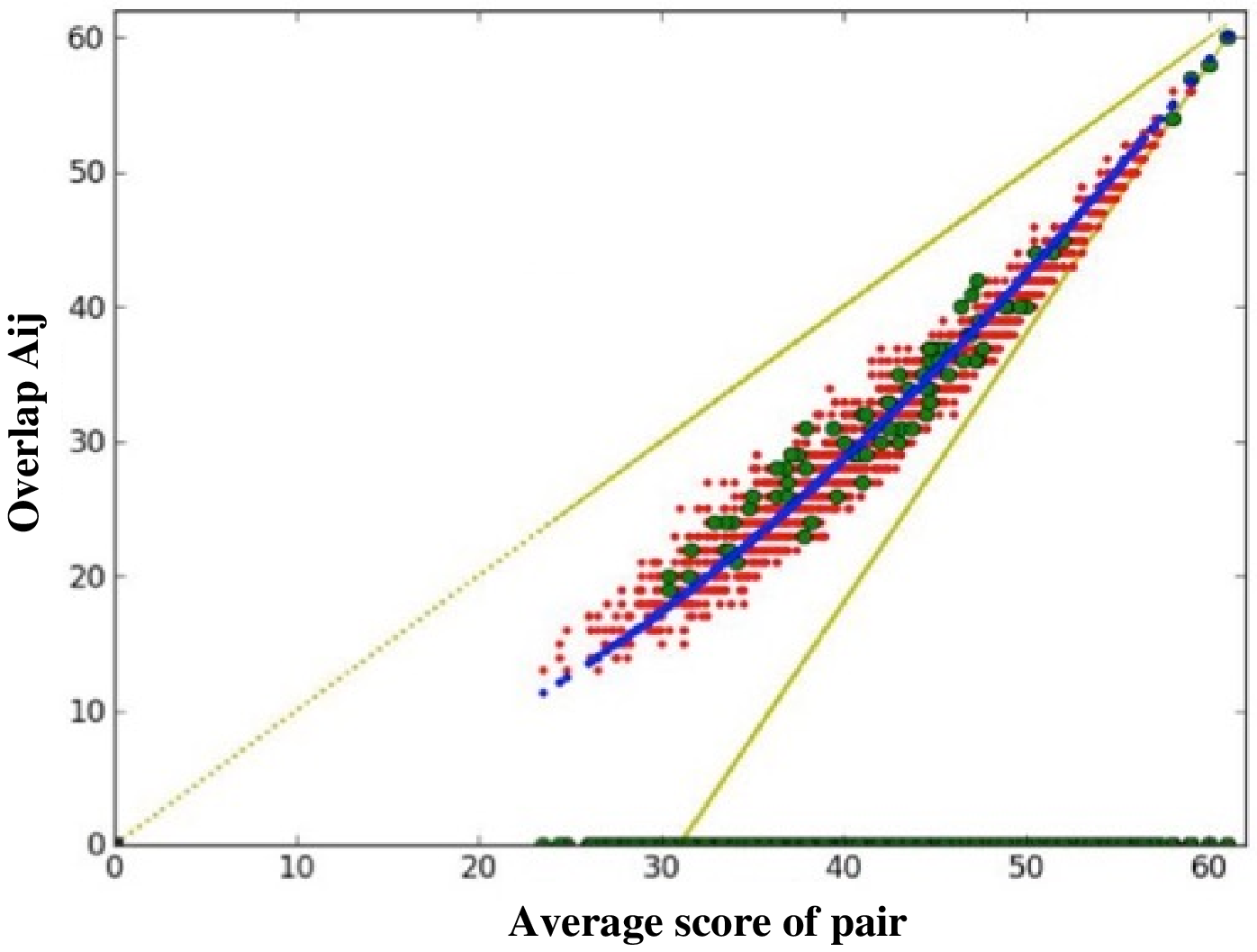}}
\qleg{Fig. 2: }
{The overlap $ A_{ij} $ as a function of the square root of the product of
  individual scores $ A_i $ and $ A_j $. 
Student pairs from the same centre are
  shown in big green dots whereas
student pairs from different centres are shown in small
 red dots. It can be observed that an electronic version of the paper
with full color figures is available on the arXiv data base of 
preprints at the following address:
http://xxx.lanl.gov/find/cond-mat.\qL}
{Source: Dataset from 2012.}
\end{figure}

\begin{figure}[htb]
\centerline{\psfig{width=12cm,figure=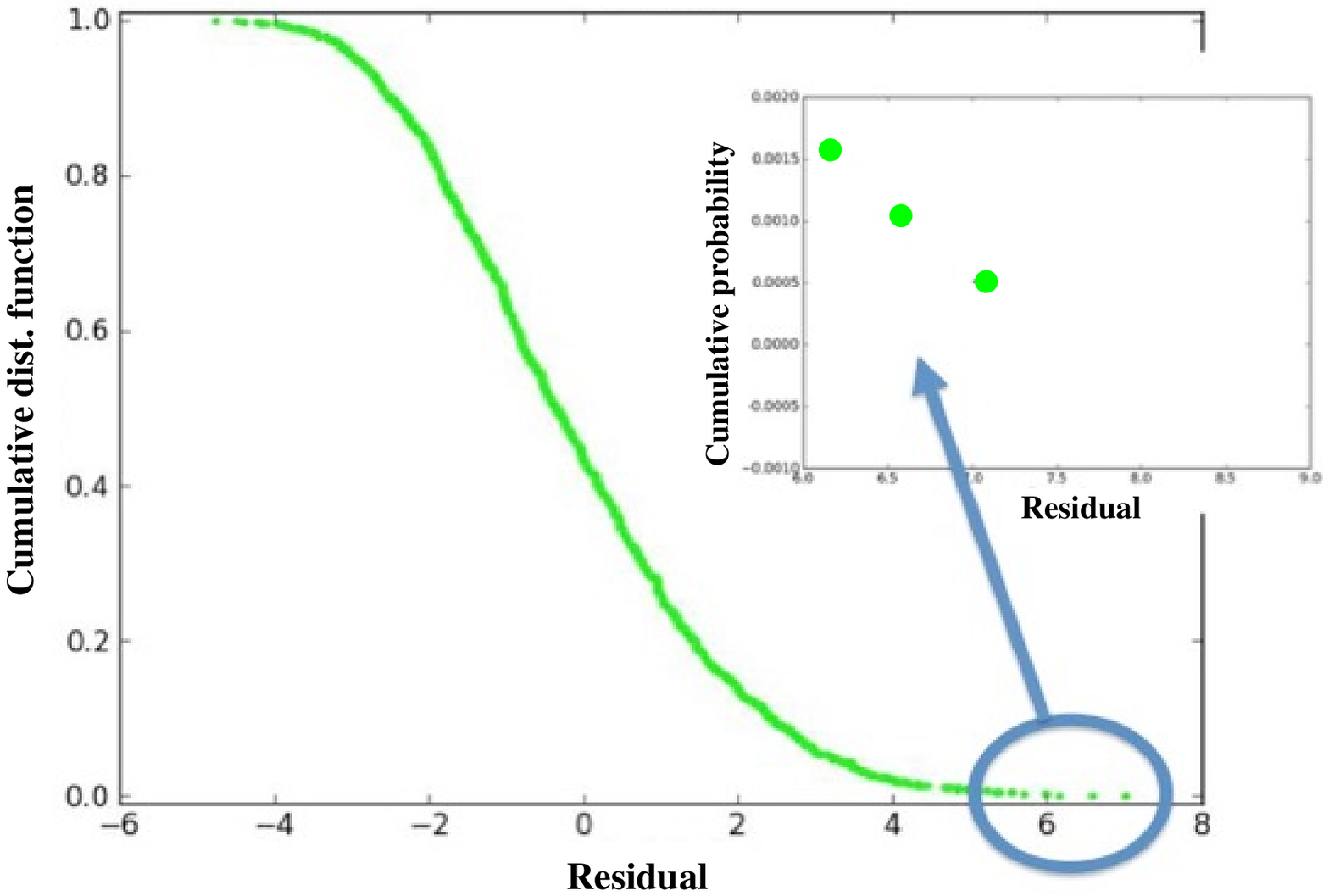}}
\qleg{Fig. 3: }
{Cumulative distribution for ``overlap'' fluctuations about the mean
  line (see text for definition) shown in figure 2. No untoward data
  points or outliers are evident from this dataset.
The total amplitude on the log scale of the vertical axis of the inset 
is from $ -0.001 $ to $ +0.002 $.}
{Source: Dataset from 2012.}
\end{figure}

However figures 4a and 4b show two further datasets from 2008 and 2011
respectively.

\begin{figure}[htb]
\centerline{\psfig{width=16cm,figure=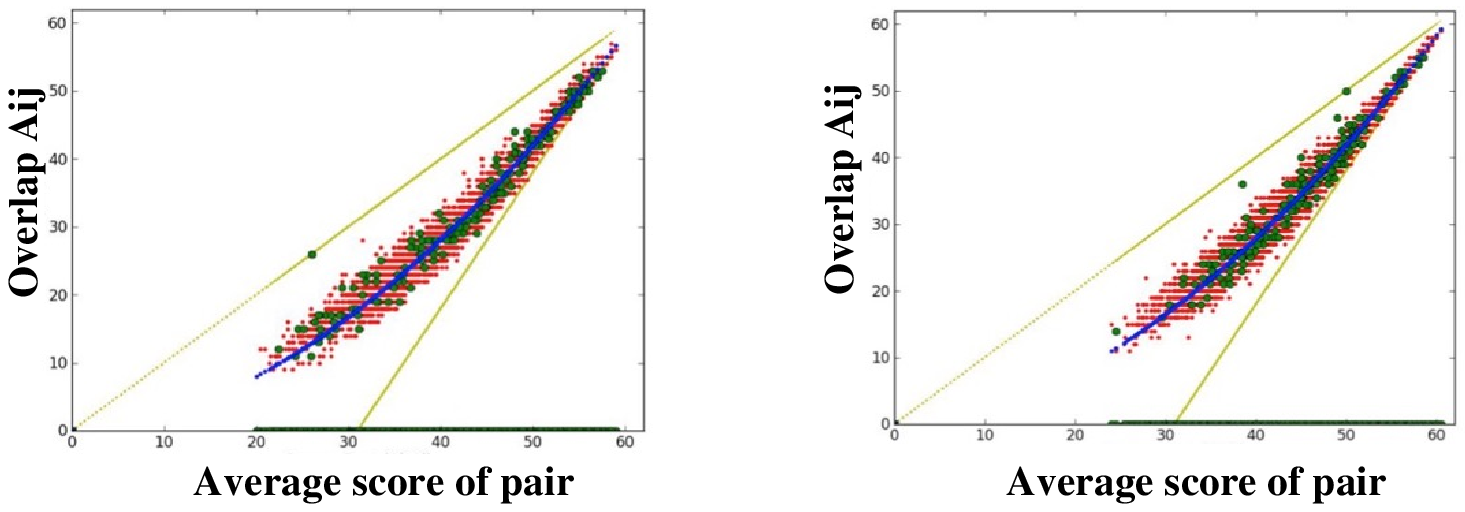}}
\qleg{Fig. 4a,b: }
{(a) Data for an examination held in 2008. (b) Similar data for an
  examination held in 2011. Both exhibit outliers that correspond to
  student pairs from the same centre.}
{Source: Dataset from 2008 and 2011.}
\end{figure}

For these examples we see a number of outliers, which become very
apparent on the corresponding cumulative distributions.

\begin{figure}[htb]
\centerline{\psfig{width=12cm,figure=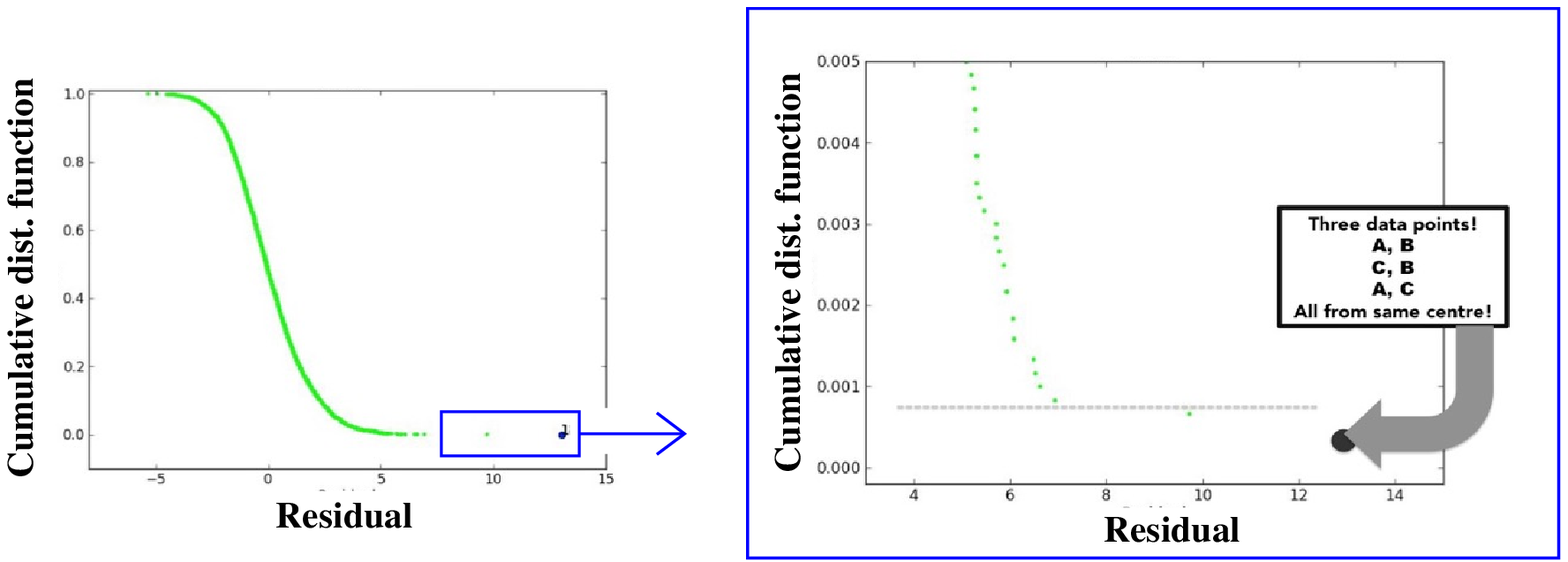}}
\qleg{Fig. 5a,b: }
{The cumulative distribution for the 2008 examination (a). Detail from
  the tail of the distribution is in figure (b) from which it now
  becomes apparent that the outlier is actually three points
  corresponding to three students from one particular centre who have
  almost identical scores and identical overlaps.}
{}
\end{figure}

\begin{figure}[htb]
\centerline{\psfig{width=12cm,figure=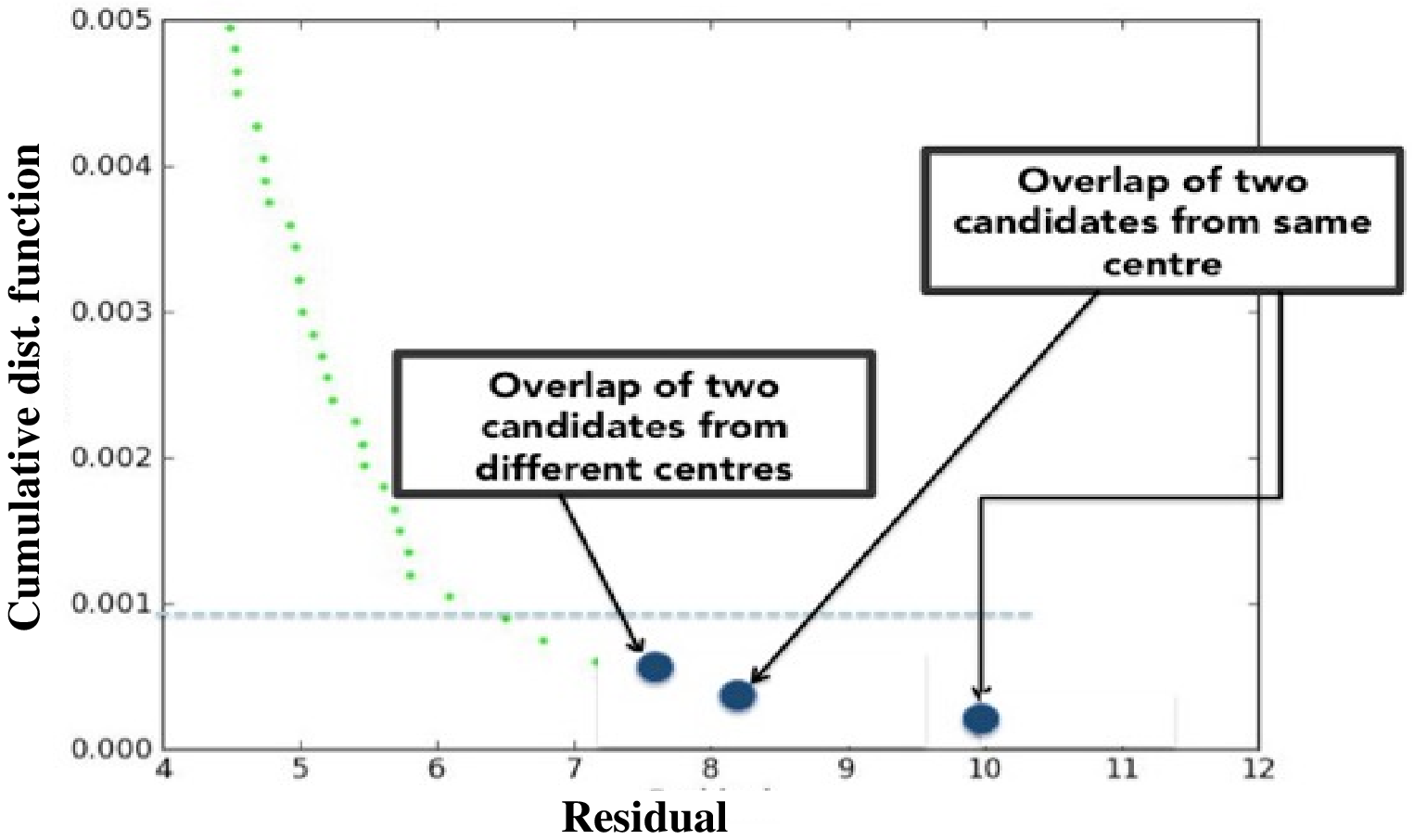}}
\qleg{Fig. 6: }
{The cumulative distribution for the 2011 examination showing two
  adjacent outliers from same centres. A small number of large random
fluctuations for candidates in different centers are
already visible in Fig. 2 and 4a,b. This is not surprising due to the
large number of pairs: $ \sim 100\times 50=5,000 $.}
{}
\end{figure}

A similar situation prevails for the 2011 examination results. However
now we see two extreme outliers, each representing a pair of students
from the same centre. This approach is similar to that described by
McManus (2005) who used it for medical examinations. However we shall now
turn to a simulation of the process, which as we shall see gives us
much greater insight into the method.

\qI{Simulation model}

It is possible to simulate situations which 
mimic the examinations. Such simulations have two main purposes.
\qbu It allows us to study the influence of major parameters
which define an examination, namely: the number of candidates,
the number of questions, the number of possible answers,
the average level of the candidates.
\qbu By giving the opportunity to run many iterations, 
the simulation allows us to judge the effectiveness of
different methods for the identification of cheating.

\qA{Principle of the simulation}
Beyond the 
technical details (which are given in the Appendix)
it is important to understand what is the nature of the probability 
problem at the core of the examination process.
\qpar

In the examinations considered in this paper there are
$ a=4 $ answers proposed for each question. Let us for a moment
suppose that there are only 2. 
Then, the whole examination process
involving $ n $ candidates and $ q $ questions 
is equivalent to throwing  $ q $ times $ n $ coins.
The face-up sides in the throw number $ k $ give the answers 
to the question number $ k $.  
In other words, the examination is a binomial random trial.
Usually, in binomial trials the coins are
supposed to be independent from one another.
If, on the contrary,
we suppose that two coins are in some way (electrically,
magnetically or in any other way) in interaction, their
results will become dependent in the same way as for the
answers of two candidates who communicate with one another.
\qpar

If we assume that $ a=6 $ the examination is equivalent to
rolling  $ q $ times $ n $ dice. As before,
the upper sides of the $ n $ dices in the trial number $ k $
will give the answers 
of the $ n $ candidates to the question number $ k $.
In other words, the examination is a multinomial random
trial. When two candidates communicate, it means that
two dice are in interaction.
\qpar

In the real examination $ a=4 $ but if we are only interested
in whether the answers are right or wrong, the examination
becomes again equivalent to a binomial trial.
\qpar

In short, the detection of cheating refers to a 
fundamental problem of probability theory, namely how to
determine whether or not two random variables are dependent.
\qpar

In
figure 7 we show the results of 62 students completing 62 questions. In
both figures it is assumed that the probability of choosing the
correct answer is constant across all questions and all students. We
don’t expect this to be a realistic assumption but we shall see that
it serves as a toy model that allows us to understand how variations
in the various aspects of the matter affect the outcomes. A feature
which is not possible using only the available empirical data. Figure
7a assumes a fairly smart set of students who answer each question
correctly with a probability of $ p=0.8 $ but the student pair, 1 and 2,
cheat with an overlap correlation of $ 0.99 $. 
Figure 7b) is similar but
the students are assumed to be less smart and the probability of
answering any question correctly is only $ p=0.5 $. The cheating pair is
immediately spotted as an outlier. The distribution of the bulk of the
data points is more limited than those for the actual student cohort
but clearly a real class will have students with different abilities
so we can expect a distribution of values for $ p $  to be more
appropriate. What is evident from the two figures is the deviation of
the data from the right hand line as the value of $ p $ 
is reduced in line
with actual data. However further calculations to explore this issue
will be left for another study.

\begin{figure}[htb]
\centerline{\psfig{width=17cm,figure=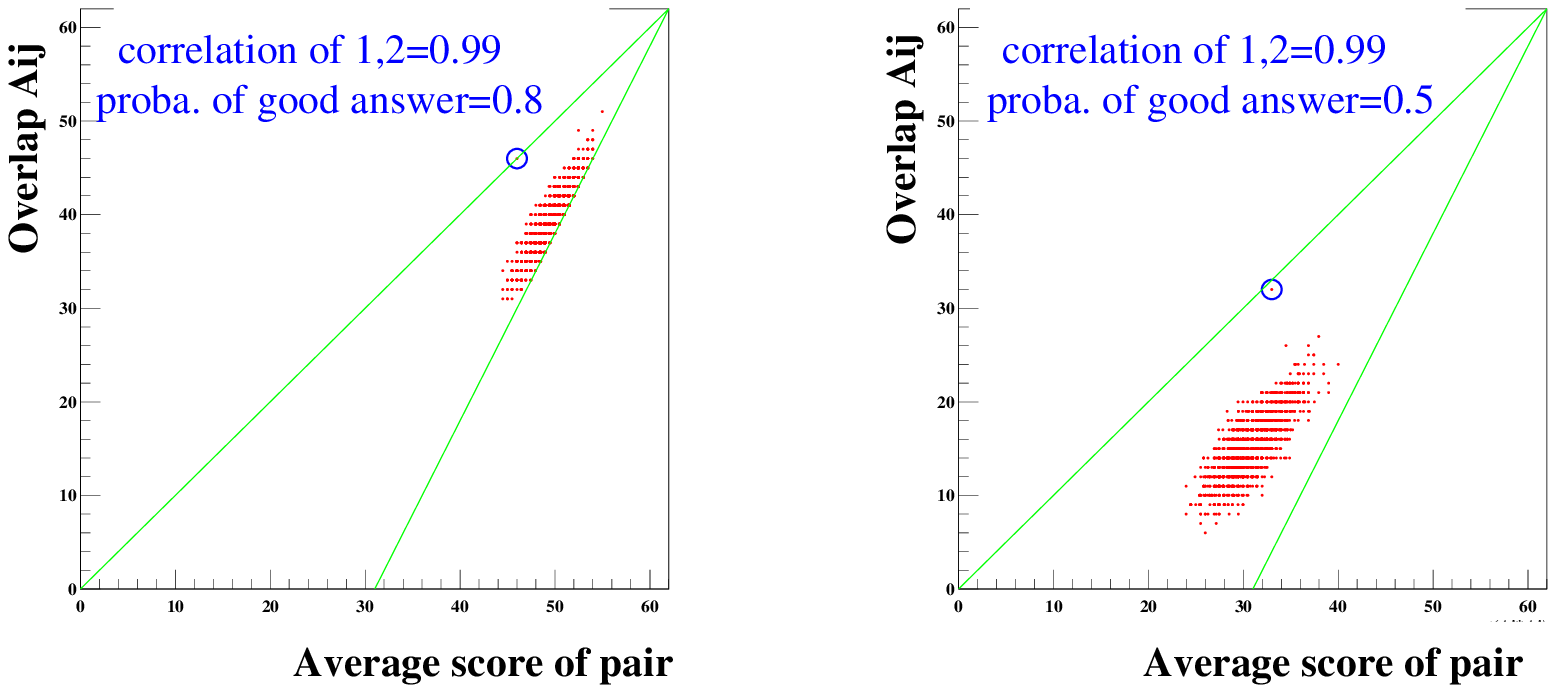}}
\qleg{Fig. 7: }
{Simulation for a group of 62 candidates. There are also 62 questions.
Each dot corresponds
to a pair of candidates. One pair, namely (1,2) is supposed
to cheat with probability 0.99 in the
  sense that the correlation between their sets of results is
  0.99. 
Figure 7a) is for candidates who answer each
  question correctly with a probability 0.8. Figure 7b) is the same
  simulation but the probability of answering questions correctly is
  only 0.5. $ A_{ij} $ is the overlap of correct answers.}
{}
\end{figure}

Figure 8 shows the degree to which correlation between the ‘cheating’
pair influences the position of the outlier.  If the correlation
parameter is less than 0.5, it would make no sense to assign it as an
outlier. But what value ought to be chosen to eliminate false
positives? We return to this point in the next section after looking
at one further outcome from the simulations. 

\begin{figure}[htb]
\centerline{\psfig{width=12cm,figure=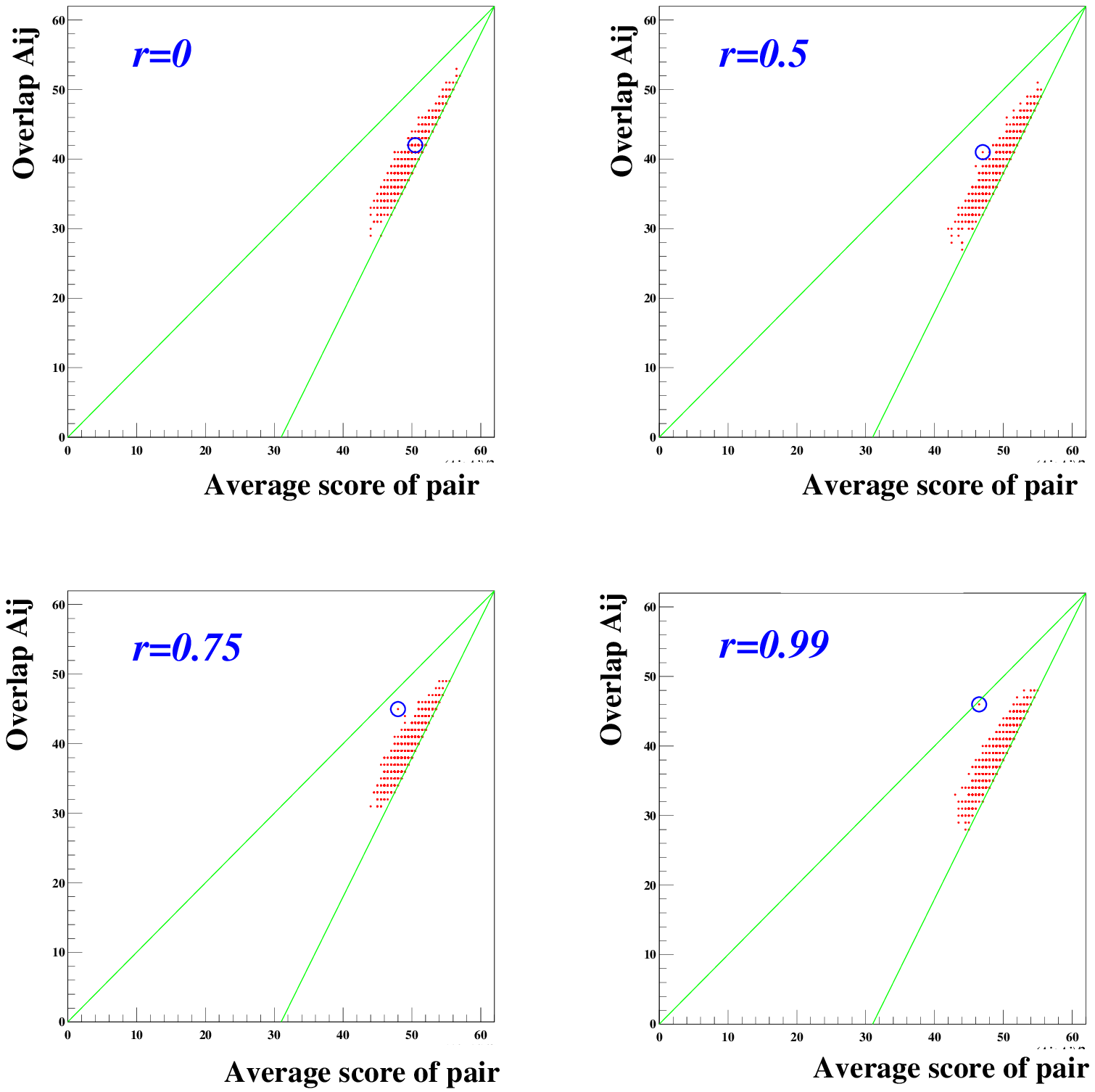}}
\qleg{Fig. 8 }
{The graphs
show the emergence of the outlier as the correlation
 between the answers of candidates 1 and 2 increases 
from zero to one.}
{}
\end{figure}

\begin{figure}[htb]
\centerline{\psfig{width=16cm,figure=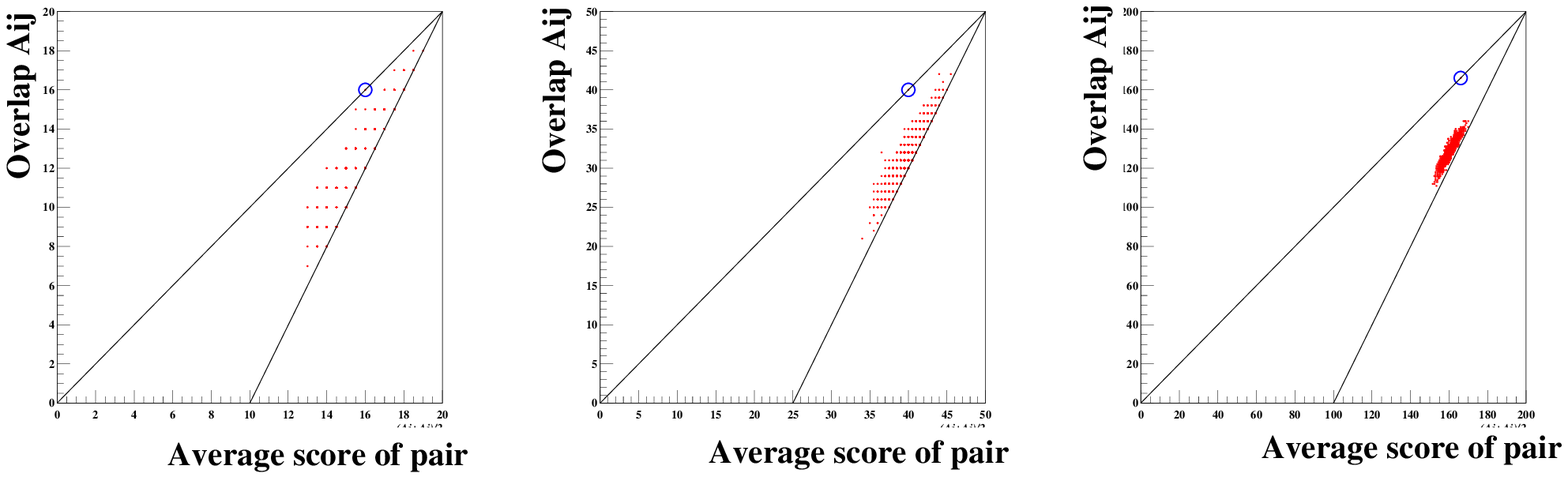}}
\qleg{Fig. 9: }
{For the 3 graphs there are 50 candidates but from left
to right the number of questions increases from 20 to 50
and 200. As expected the outlier is more obvious
when the number of questions becomes larger.}
{}
\end{figure}

Figure 9 shows the effect on the results of changing the number of
students relative to the number of questions in the examination
paper. We see that an outlier is more obvious when the number of
students is less than the number of questions. As an aside we note
here that the total number of student pairs remains constant across
the graphs; that this may not appear so reflects the fact that many
points are superimposed when
integer values of the overlap coincide.

From figure 10, we see that an outlier is also more obvious for
students who are not so smart than when they all score highly. 

\begin{figure}[htb]
\centerline{\psfig{width=16cm,figure=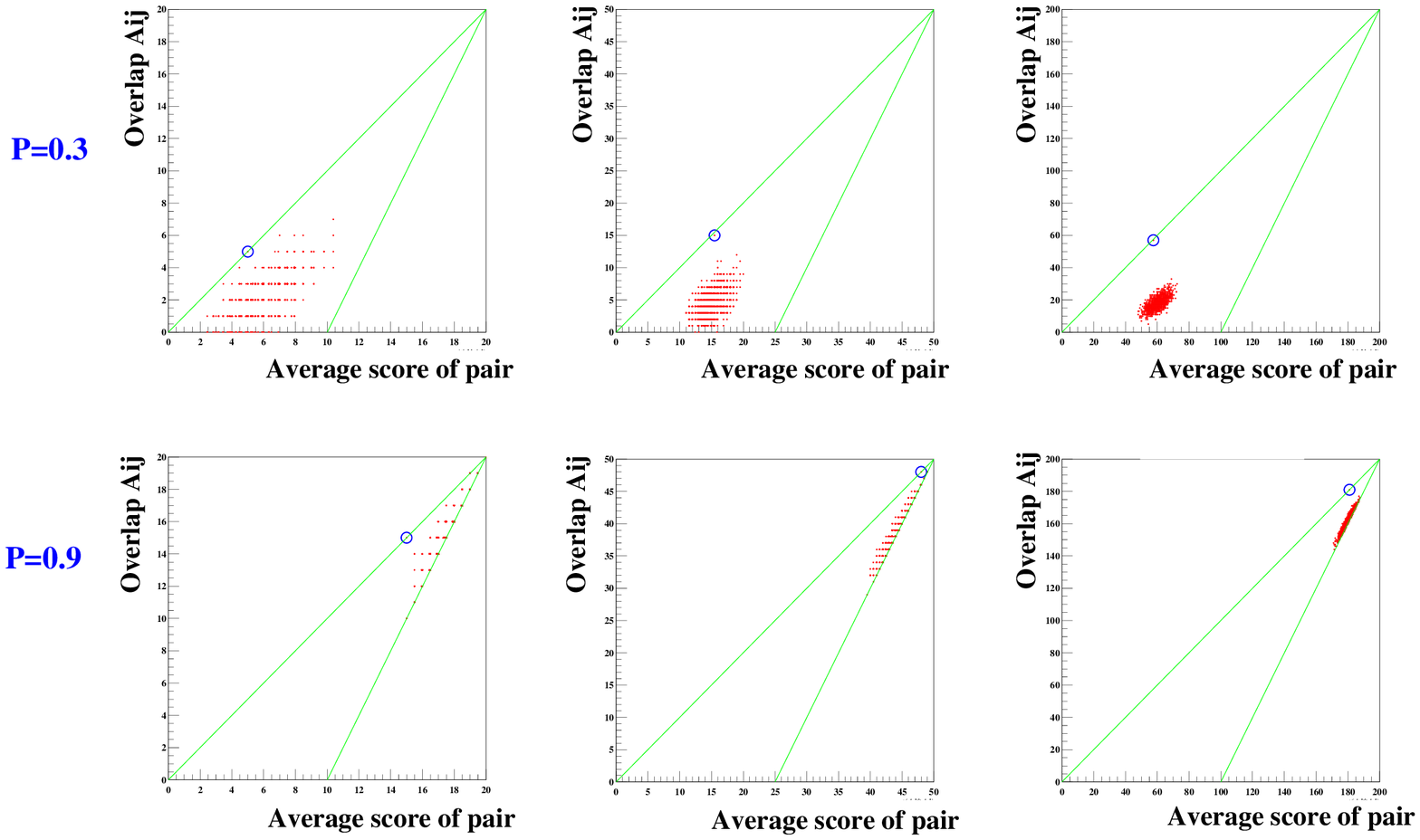}}
\qleg{Fig. 10: }
{These graphs combine two effects: change in the smartness of the
students through the probability $ p $ and change in the number
of questions. The two effects work in opposite directions. The
bottom right graph shows that when $ p=0.9 $ even with
a number of questions as high as 200
(which from a practical perspective seems to be an upper limit) 
the outlier is hardly detectable.}
{}
\end{figure}

Our examination data is usually for a cohort which is greater in
number than 62, the number of examination questions. But they are not
all equally smart and moreover our network, as we noted at the outset
is fragmented, with many centres hosting only a few
examinees. Overlaying the sub-network on the complete network of
overlaps as we showed earlier now helps overcome any potential masking
of the outliers. Examples are shown in Figure 4a,b.
\qpar

The majority of data points for the sub-network are more confined
forming a slightly tighter distribution within the total network
leaving the single outlier in this example more clearly exposed. As
for the 2012 data the majority of points within the sub-network
seem are more or less randomly scattered about the mean relative to
the totality of points, so again we cannot conclude that students who
study together gain any additional advantage over those who do not and
once more we see that the frequent assertion that students who study
together will have common overlaps does not pass muster.
\qpar

Wesolowsky1 stresses the importance of using a conservative method to
identify false positives. He kindly agreed to process our data using
his proprietary method and the specific ‘cheaters’ he identified are
the same as obtained here. He imposes a boundary value of 
$ 10^{-4} $ on the
probabilities shown in the cumulative distributions in figures 5b and
6 as a light blue line. Points lying below this value are the
outliers. We may understand this by plotting our data on a log-log
scale and superimposing a Gaussian distribution chosen to fit the
points in the main bulk of the distribution.  Figure 11 illustrates
the detail. The green line is the cumulative distribution function for
our data; the red line is a Gaussian fit to the bulk of the data. In
this case the outlier has a probability 8 orders of magnitude greater
than that which we would expect from the Gaussian distribution
function. Simple enough to identify one would think. But at this point
it is usual to apply a statistical test ascribed to Bonferroni in order
to reduce the chance of false positive results%
\qfoot{For more details see the Wikipedia article
entitled `` Bonferroni correction''; in particular
see the references to the statistical papers cited 
in this article.}%
.
It is argued that the
distribution function computed from the data represents the outcome of
a single experiment. However we have N candidates sitting the
examination. The Bonferroni correction says that the true probabilities
ought to be  $ N(N-1)/2 $ greater than the single Gaussian values. It is
not easy to understand why this should be so. Moreover in our case the
application of this is a little ambiguous since candidates take the
examination in small groups. Do we choose a value of the order of the
group size typically 3 to 6 or do we choose $ N=63 $ the total size of the
cohort? We shall ignore this and select the most conservative approach
which is to choose $ N=63 $.  
Our correction factor is then of the order
of $ 10^{-4} $ which means the corrected probability is still far less than
that observed in this case so we can be quite sure that our outliers
are true outliers and not just erroneous points associated with the
Gaussian.
\qpar

\begin{figure}[htb]
\centerline{\psfig{width=12cm,figure=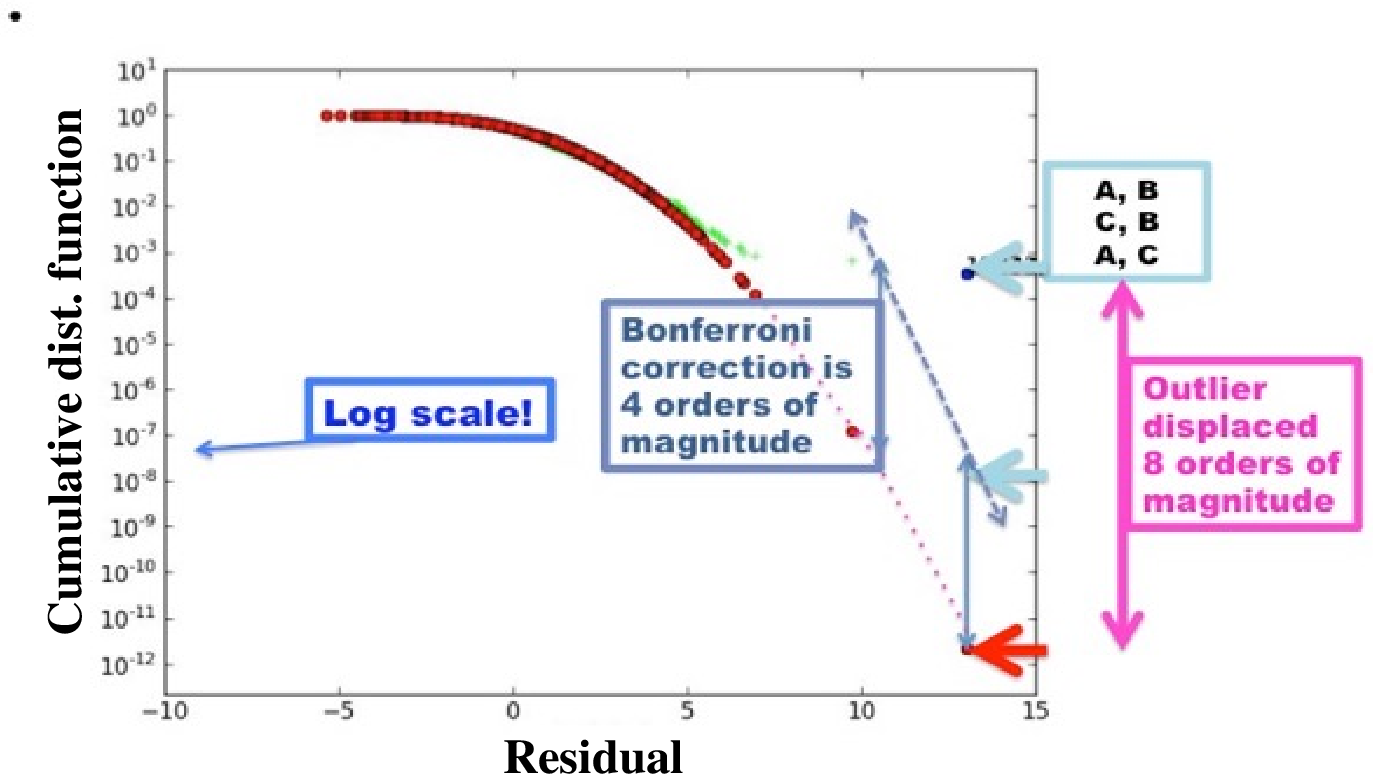}}
\qleg{Fig. 11: }
{The cumulative distribution of overlaps for the data of 2008
plotted on a log log scale. Now the deviation of the actual
distribution (circles) from a Gaussian (points) is clear. The three
data points from the same examination centre and suggestive of
collusion are visible and well separated from the Gaussian curve. The
Bonferroni correction offers only 4 orders of magnitude of hope leaving
a further 4 orders of magnitude unaccounted for! }
{}
\end{figure}

Analysis of the second data (figure 6) set suggests both outliers are
true examples of anomalous pairs each being for students from the same
examination centre. Although the penultimate point only just falls
within this conservative cut-off!

\qI{Overlaps or correlations?}

The method of looking at overlap of correct answers 
(as used by McManus et al. 2005) does not use
information relating to wrong answers. Moreover, and this point
is probably more meaningful, it does not take into the specific
answer numbers selected by the candidates.
In this
section we propose an alternative method based on the correlations
between the two results vectors of pairs of candidates. 
In the present case the results vector of a candidate
is a set of 62 numbers comprised between 1 and 4.
\qpar

Why should one use correlations when in this problem the
natural variable is the overlap between the answers of different
candidates? When a teacher begins to suspect some
``abnormalities'', the first reaction is to assess 
the similarity in the
answers of the candidates. That is why the overlap was used as a
key-variable in the first part of this paper.
\qpar

However, the overlap is not a standard statistical concept and does
not easily lead to an assessment in terms of confidence
intervals. That is why, instead of the overlap, we will here use the
correlation. Overlap and correlation are two measures of the degree of
similarity between two sets of results and it is easy to check that
the two variables increase together when the results become more
identical. However by using correlations we can use standard theorems
that give confidence interval for a given level of confidence
probability. 
\qpar

More specifically the confidence interval limits that we will use are
obtained following Morice and Chartier (1954). 
Lying between $ -1 $ and $ 1 $, the
correlations can certainly not be considered as Gaussian
variables. That is most unfortunate because the confidence intervals
of Gaussian variables can be obtained most easily. Therefore it is
natural to apply a change of variable which will transform the interval
$ (-1,1) $ into an interval covering the whole line from minus infinity to
plus infinity. The inverse hyperbolic tangent function does this.
Then
the correlations so transformed can be fairly well described by a
Gaussian variable. One we have performed
the standard
confidence analysis for Gaussian variables we return to the
correlations by way of applying an hyperbolic tangent function. 
\qpar

Before using this method
we need to decide the kind of results to which we want to apply
the methodology. There are basically two possible options.
\qpar

\begin{figure}[htb]
\centerline{\psfig{width=14cm,figure=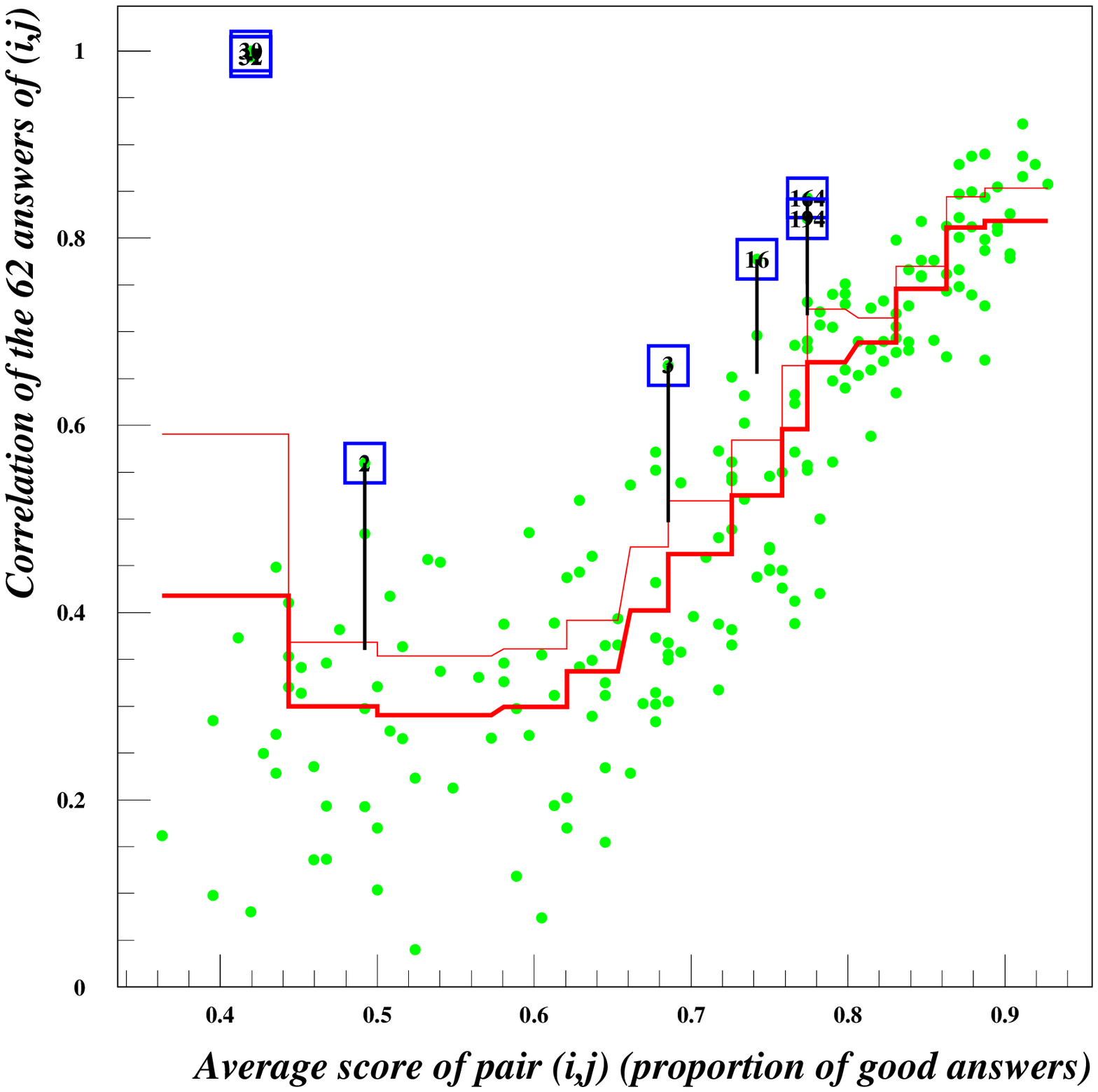}}
\qleg{Fig. 12a: Analysis of 2008 data (110 candidates) in terms of
correlation between the answers to the 62 questions.}
{ Answers are
described as vectors whose elements are numbers comprised between 1
and 4 depending on which one
of the 4 possible answers the candidate selected. 
The thick (red) staircase curve
gives the averages for groups of 14 successive points.
The thin line gives the confidence interval of the averages
(at probability level of 0.95).\qL
It can be observed that the correlations $ C_{ij} $ of the answers 
used in this graph
are closely related to their overlap in terms of
right/wrong answers, that is to say the $ A_{ij} $ used in earlier graphs;
the correlation between the $ C_{ij} $ and $ A_{ij} $ is $ r=0.97 $.\qL
The outliers whose confidence intervals (also with probability
level of 0.95 and represented by the 
black vertical lines) do not cross the staircase curve
can be considered as having a significantly abnormal correlation.\qL
Altogether there are 8 such pairs: the most conspicuous
correspond to the three squares at the top of the graph
almost at probability 1 (its confidence interval is so small
that it does not appear on the graph).
They correspond to the three candidates who took the examination at
the same centre and identified already in Fig. 5 by the Gaussian
method.
The 5 other squares, namely those numbered 2, 3, 16, 164 and 194,
correspond to 5 pairs of candidates who took their
examination in 4 centres $ C_1,C_2,C_3,C_4 $. We can
say that pair number 2 and pair number 3 took their
examination in the same centre $ C_1 $, whereas the 3 other centres
were all different. The exact code numbers of the centres
as well as the code numbers of the candidates have been removed
for confidentiality reasons.}
{Source: Data set from the UK Radio Communication Foundation
(2008).}
\end{figure}

\qpar

\begin{figure}[htb]
\centerline{\psfig{width=14cm,figure=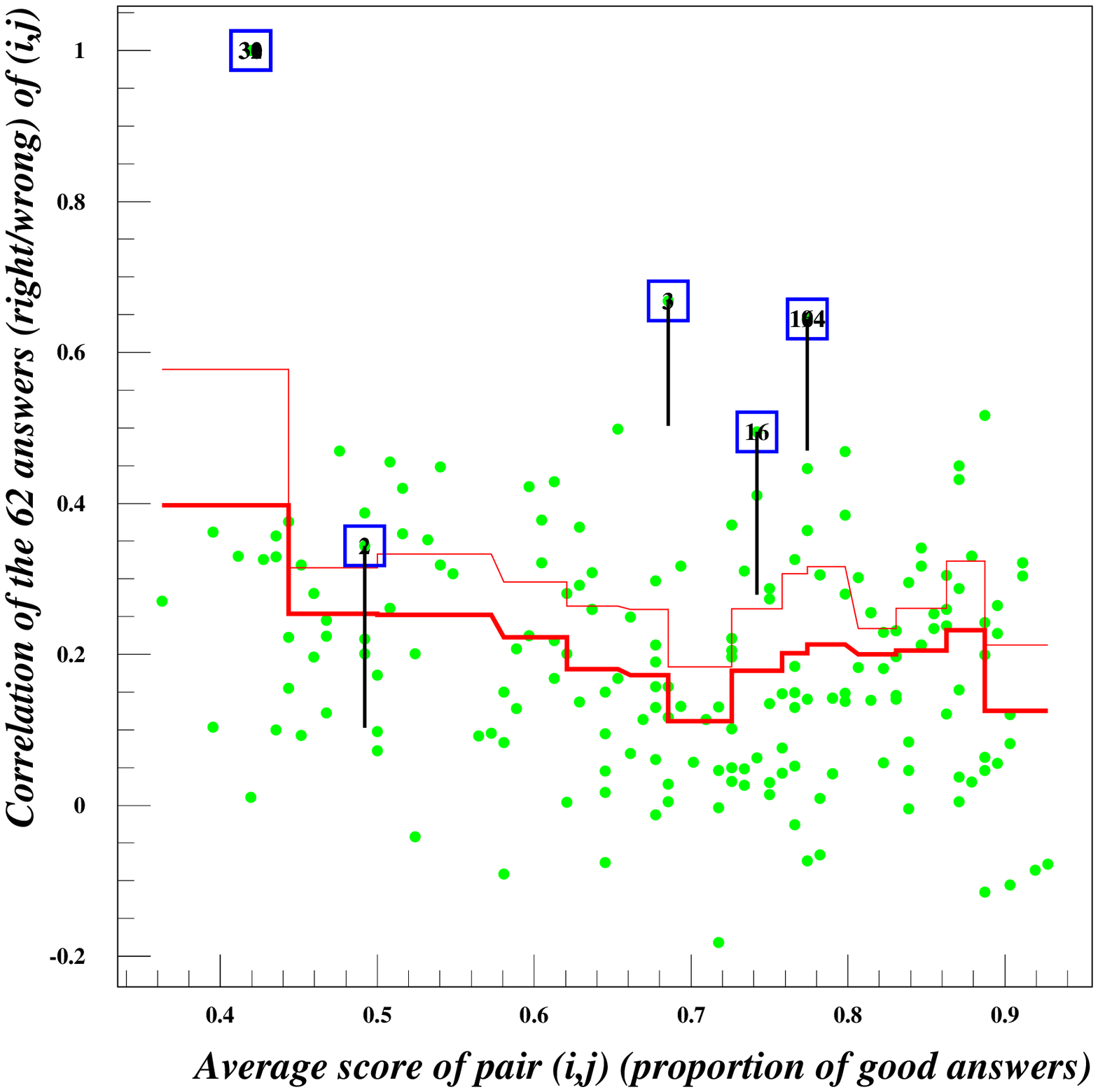}}
\qleg{Fig. 12b: Analysis of 2008 data (110 candidates) in terms of
correlation between the  right/wrong characteristics of the 62
answers.}
{Answers are
described as vectors whose elements are 1 or 0 depending on
whether the answer is right or wrong.
The staircase curves have the same meaning as in Fig. 12a.
It can be observed that the correlations $ C_{ij} $ of the answers 
used in this graph
are less closely related to their overlaps $ A_{ij} $; 
the correlation is $ r=0.75 $. This is mainly due to the fact that
the $ C_{ij} $ are lower and hence more dispersed (due to
larger confidence intervals) than in Fig. 12a. \qL
With the exception of the data point number 2,
the outliers are the same as in Fig. 12a. In contrast
with Fig. 12a, the correlation does not display an upward
trend. This is due to the fact that for low scores the vectors
of the answers contain many 0 in the same way as they have
many 1 for high scores; this makes the situation fairly
symmetrical; it can be noted that 
the high level of the first step of the 
staircase curve is due to the inclusion into the average of the
3 top data points with correlations almost equal to 1.}
{Source: Data set from the UK Radio Communication Foundation
(2008).}
\end{figure}

The first approach is to use the results in the form ``good answer''
versus ``wrong answer''. In this framework, the set of results of each
candidate will be a simplified vector whose elements are 0 (wrong
answer) or 1 (correct answer). For the sake of simplicity, the
previous simulations have been performed using this framework.
\qpar

Alternatively we can use all the available information. As every
teacher knows, common wrong answers are more suggestive of possible
cheating than common good answers. For instance, if the correct answer
is A, two D answers may mean either that the two candidates had the
same misunderstanding or that there was some form of communication
between them. In the present case where there are 4 possible answers
the set of results of each candidate will be a vector whose elements
can take the values 1,2,3,4%
\qfoot{As can easily be checked, the correlation between two sets
of answers remains the same if instead of 1,2,3,4 we use 11,12,13,14
or 1,3,5,7. Whereas the first case is obvious because it is
merely a translation, the second is somewhat less obvious.}%
.
Because it uses more information it is of course
natural to expect that this second approach will lead to more accurate
identification of interdependent results. This is indeed confirmed by
observation as shown in Fig. 12a,b.
\qpar

As in previous graphs, green dots correspond to pairs of candidates
who took their examination in the same centre. The black vertical lines
shown for a number of outliers are the lower parts of the confidence
intervals for a probability level of 0.95 (which means that if we
repeat 100 times the same experiment about 95 of the correlations for
the same pair of candidates will fall in the confidence interval). For the 3
candidates which correspond to the two squares at the top (almost at
correlation 1) the confidence interval is so small that the 
confidence interval
can hardly be seen. This is of course an
extreme case in the sense that 99\% of the answers selected by these 3
candidates were identical. In such a case cheating is hardly in doubt.   
\qpar

However, for the other cases the situation is less clear. If we wish
to interpret them in terms of cheating or not cheating an additional
assumption is necessary, namely one needs to assume that the bulk of
the candidates did {\it cheat}. This assumption has already been made in
the Bonferroni method. In cases where all candidates would cheat the
identification of outliers would only reveal the candidates who cheat
more than the average. 
\qpar

With this assumption in mind, one can adopt the rule that when the
confidence interval does not cross the staircase line that
represents the average of all candidates, there is a significant
``abnormal'' similarity. Needless to say, such a statement depends on
the confidence level that is selected. With a confidence level of 0.99
(instead of 0.95), the confidence intervals would be 
broader and the number of
significant cases would be reduced.

\qI{Conclusions}

\qA{How can one optimize the detection of cheating?}

The method of overlaps, that compares correct answers for pairs of
candidates, is shown to be effective at identifying anomalous outcomes
suggestive of cheating, in multiple-choice examinations. Simulation
confirms the method and offers insight into parameters that might be
controlled in future examinations to optimize such identification. The
method can be criticized for focusing on overlap only of correct
answers, taking no notice of overlap of incorrect answers. Intuitively
it seems that overlap of incorrect answers could be a stronger pointer
to collusion. The use of correlations allows the use of well-defined
and more widely understood confidence limits for eliminating false
positives, rather than the Bonferroni method which becomes rather
ill-defined in our case where the examination cohort is split across
numerous centres. Since it is conceivable that cases of cheating could
be the subject of legal action, it is important that well understood
methods for elimination of false positives are used. We show that the
method of correlations appears to be equally applicable in our case
giving identical outcomes. 
\qpar
A practical way to minimize the potential for cheating in multiple
choice examinations such as those discussed here is to randomize the
order of potential answers on candidate question sheets and this
approach is now used by the RCF in the UK. 

\qA{Extension of the present analysis to similar problems}

This paper is about a specific problem, namely
how to detect people who cheat at examinations.
However, we believe that the identification
methods proposed in the paper can also be used
every time one wants to detect ``abnormal''
similarities in a selection process.
\qpar
This can be illustrated by the following example.
We consider a group of $ N $ customers.
Every customer is shown 4 brands of various products
(tea, milk, beer, rice, cookies and so on) and is asked 
which one he or she prefers.\qL
The dispersion of the correlations (or overlaps) 
of the vectors of selected brands
will provide a way to measure how similar or
closely connected the customers are.
The analog of cheating will correspond to
an abnormally high correlation suggesting
a strong connection between two customers.
\qpar

One can also introduce the notion of right/wrong answers.
For instance, ``right'' answers may
be defined as the brands for which there had been a recent
advertisement campaign. 
Then, one can define the score of a customer as the number
of his (or her) correct answers. The customers with highest
scores (the analogue of the smart students) are likely to
be those who watch TV the most.

\vskip 3mm

{\bf Acknowledgments}
We are grateful to the Radio Communications Foundation of the United
Kingdom for permission to use their data. This collaboration began
with a discussion at ‘Ways of Seeing’, a workshop sponsored by COST
action TD1210 Analyzing the dynamics of information and knowledge
landscapes - KNOWeSCAPE and held at the University of Galway during
April 2014.  The authors are grateful to the action chair and
management committee of the COST action for their support.\qL
The authors wish to thank their referees for their careful reading
and insightful suggestions.

\appendix

\qI{Appendix: Simulating interdependent random variables}

In this appendix we wish to answer two related questions
regarding the definition of a system $ X_i $ of 
symmetric binary interdependent
random variables:
$$ x_i\in \{0,1\} \quad E(X_i)=p \quad \sigma^2(X_i)=p(1-p)\quad
\hbox{corr}(X_i,X_j)=\rho_{ij}=r\quad 1\le i,j\le n \qn{1} $$
\vskip -5mm
\qee{1} In the simulation we used a system $ \{X_i\} $
introduced by Lunn and Davis (1998). How is it
defined?
\qee{2} Apart from the previous one,
are there other $ \{X_i\} $ systems which share the properties 
defined in (1)? 
\qL
It will be seen that it is only
for $ n=2 $ that the correlation matrix $ \rho_{ij} $
 {\it uniquely} defines the probability distribution. 
In other words, the set of random variables introduced by
Lunn and Davies is only one among many possible answers.

\qA{The set  $ \{X_i\} $ proposed by Lunn and Davies (1998)}

It is based on the following result. $ (0,1;p) $ denotes a binary
variable whose expectation is $ p $.
\qdec{
{\it Proposition} The $ Y_i, (1\le i\le n $) are 
a set of independent binary
random variables $ (0,1;p) $. $ Z $ is also a binary variable 
$ (0,1;p) $. The $ U_i $ are a set of binary variables $ (0,1;\sqrt{r}) $.
The $ Y_i,Z,U_i $ are all independent.\qL
Then the variables $ X_i=(1-U_i)Y_i+U_iZ $ are a system of 
symmetric dependent 
binary variables $ (0,1;p) $ whose correlation matrix is
$ \rho_{ij}=r $.
}
\qpar
The proof is easy and given in the paper. Let us add two
remarks.
\qbu We used the expression ``symmetric variables'' to reflect
the fact that all $ X_i $ play the same role. The expression
``exchangeable variables'' is often used with the same meaning.
\qbu The correlation matrix has only positive elements.
This is of course imposed by the symmetry condition. 
$ \rho_{12}<0 $ and $ \rho_{23}<0 $ would imply $ \rho_{13}>0 $,
thus violating the symmetry requirement.

In the following subsections we will be concerned with the question of
uniqueness of the set of $ X_i $ generated above.
Needless to
say, it is useful to know whether the Proposition gives {\it the} answer 
or only one among many. More precisely, the problem can be stated as
follows. 
\qdec{In order to specify the probability distribution of one 
$ X_i $ variable one needs $ h $ parameters. In addition,
the correlation matrix contains one parameter. Are these $ h_1=h+1 $
parameters sufficient to determine the joint probability distribution
of the whole set $ \{X_i,\ i=1,\ldots ,n\} $.}

\qA{Uniqueness for two binary dependent variables}

The definition (1) contains only two parameters, namely $ p $
and $ r $. In this case $ h_1=2 $.
It will be seen that many systems (indeed an infinite number)
of variables can be defined which fulfill these conditions.
The only exception is the case $ n=2 $. In this case the correlation
matrix defines completely the probability distribution.
\qpar
For two variables $ X_1,X_2 $ the probability distribution function
is completely defined by two joint probabilities:
$$ p_{11}=P\{X_1=1 \hbox{ and } X_2=1\},\  
 p_{01}=p_{10}=P\{X_1=1 \hbox{ and } X_2=0\}\quad \hbox{then: }
p_{00}=1-p_{11}-2p_{01} $$

Thus, through an argument based on degrees of freedom, one sees
that there is a one-to-one correspondence between the parameters
$ p,r $ and the probabilities $ p_{11},p_{01} $. 

\qA{Non uniqueness for more than two binary variables}

On the contrary, a system of
3 variables is defined by 3 probabilities, namely
$$ p_{111}, p_{011}, p_{001},\quad \hbox{ with: } 
p_{000}=1- p_{111}-3p_{011}-3p_{001} $$

Thus, one of these 3 numbers can be chosen arbitrarily.
There is no longer a one-to-one correspondence with the parameters
$ p,r $.
\qpar

The same property can be seen in a slightly different way.\qL
Clearly, $ E(X_1X_2)=1\times 1\times p_{11} $; similarly
$ p_{01}=E[(1-X_1)X_2]=p-E(X_1X_2) $. This shows that the
probabilities which define the distribution are 
specified by $ E(X_1X_2) $ that is to say basically by the
correlation matrix. As a matter of fact, an easy calculation
gives the expressions of the probabilities ($ q=1-p $):
$$ p_{11}=r^2pq+p^2,\quad p_{01}=p_{10}=pq(1-r^2),\quad
p_{00}=r^2pq+q^2 $$

On the contrary, the expression of $ p_{111} $, namely
$  p_{111}=E(X_1X_2X_3) $, shows that for 
3 binary variables one needs a three-point moment. The 
two-point moment of the correlation matrix 
is no longer sufficient.

\qA{Non uniqueness for all non-binary variables}

The uniqueness property for the two-variable case
is special to binary variables. 
\qpar
If $ X_i $ can take 3 values, say $ 0,1,2 $, its probability
function is defined by two numbers, for instance $ p_0,p_1 $,
with $ p_2=1-p_0-p_1 $. Thus, $ h_1=3 $. However, for
two variables one needs 4 numbers to define the symmetric joint 
probability distribution namely: 
$$ p_{00},p_{10},p_{20},p_{11},\quad \hbox{ with: } 
p_{22}=1-p_{00}-2p_{10}-2p_{20}  $$

Thus, there is no uniqueness.

\vskip 5mm

{\bf References}

\qparr
Cizek, G.J. (1999). Cheating on tests: how to do it, detect it, and
prevent it. Mahwah (New Jersey): Lawrence Erlbaum.

\qparr
Eliana Dockterman, Time Magazine, May 8 2014.

\qparr
Lunn (A.D.), Davies (S.J.) 1998: A note on generating
correlated binary variables. Biometrika 85,2,487-490.

\qparr
McCabe D.L, (2005) Cheating among college and university students: A
North American perspective. International Journal for Educational Integrity,
Vol. 1, 1.

\qparr
McManus, I.C., Lissauer T., S E Williams S.E. (2005): Detecting
 cheating in written medical examinations by statistical analysis of
 similarity of answers: pilot study.
 British Medical Journal 330:1064–1066.

\qparr
Morice E., Chartier F (1954) M\'ethode statistique. Part 2: Analyse
statistique, Institut National de la Statistique et des Etudes
Economiques (INSEE), Paris, National Printing Office, page 371.

\qparr
Wesolowsky G.O. (2000) Detecting excessive similarity in answers on
multiple choice exams. Journal of Applied Statistics, Vol. 27,
909-921.

\end{document}